\UseRawInputEncoding

\documentclass[journal,transmag]{IEEEtran}
\usepackage{siunitx}
\usepackage[german]{babel}

\usepackage{textgreek}

\ifCLASSINFOpdf
   \usepackage[pdftex]{graphicx}
   \graphicspath{{../pdf/}{../jpeg/}}
   \DeclareGraphicsExtensions{.pdf,.jpeg,.png}
\else
   \usepackage[dvips]{graphicx}
   \graphicspath{{../eps/}}
   \DeclareGraphicsExtensions{.eps}
\fi
\begin{document}
%
\title{Temperature-dependent spectral properties of (GaIn)As/Ga(AsSb)/(GaIn)As ``W''-quantum well heterostructure lasers}


\author{\IEEEauthorblockN{Christian~Fuchs\IEEEauthorrefmark{1,2},
Ada~B\"aumner\IEEEauthorrefmark{1},
Anja Br\"uggemann\IEEEauthorrefmark{1},
Christian Berger\IEEEauthorrefmark{1},
Christoph M\"oller\IEEEauthorrefmark{1},
Stefan~Reinhard\IEEEauthorrefmark{1},\\
J\"org~Hader\IEEEauthorrefmark{3,4},
Jerome~V.~Moloney\IEEEauthorrefmark{3,4},
Stephan~W.~Koch\IEEEauthorrefmark{1},
Wolfgang~Stolz\IEEEauthorrefmark{1,2}}
\IEEEauthorblockA{\IEEEauthorrefmark{1}Materials Sciences Center and Department of Physics, Philipps-Universit\"at Marburg, Renthof 5, 35032 Marburg, Germany.}
\IEEEauthorblockA{\IEEEauthorrefmark{2}NAsP\textsubscript{III/V} GmbH, Hans-Meerwein-Straße, 35032 Marburg, Germany.}
\IEEEauthorblockA{\IEEEauthorrefmark{3}College of Optical Sciences, University of Arizona, 1630 E. University Blvd., Tucson, AZ, 85721, USA.}
\IEEEauthorblockA{\IEEEauthorrefmark{4}Nonlinear Control Strategies Inc., 7562 N. Palm Circle, Tucson, AZ, 85704, USA.}
\thanks{Manuscript received November XX, 2019; revised December YY, 2019.
Corresponding author: C. Fuchs (email: christian.fuchs@physik.uni-marburg.de).}}


\markboth{IEEE Journal of Quantum Electronics,~Vol.~XX, No.~YY, December~2019}%
{Fuchs \MakeLowercase{\textit{et al.}}: Temperature-dependent spectral properties of (GaIn)As/Ga(AsSb)/(GaIn)As ``W''-quantum well lasers}
%



\IEEEtitleabstractindextext{%
\begin{abstract}
This paper discusses the temperature-dependent properties of (GaIn)As/Ga(AsSb)/(GaIn)As ``W''-quantum well heterostructures for laser applications based on theoretical modeling as well as experimental findings.
A microscopic theory is applied to discuss band bending effects giving rise to the characteristic blue shift with increasing charge carrier density observed in \mbox{type-II} heterostructures.
Furthermore, gain spectra for a ``W''-quantum well heterostructure are calculated up to high charge carrier densities.
At these high charge carrier densities, the interplay between multiple \mbox{type-II} transitions results in broad and flat gain spectra with a spectral width of approximately \SI{160}{\nano\metre}.
Furthermore, the temperature-dependent properties of broad-area edge-emitting lasers are analyzed using electroluminescence as well as laser characteristic measurements.
A first indication for the theoretically predicted broad gain spectra is presented and the interplay between the temperature-dependent red shift and the charge carrier density-dependent blue shift is discussed.
A combination of these effects results in a significant reduction of the temperature-induced red shift of the emission wavelengths and even negative shift rates of (-0.10 $\pm$ 0.04)\,\si[per-mode=symbol]{\nano\metre\per\kelvin} are achieved.
\end{abstract}

}

\maketitle

\IEEEdisplaynontitleabstractindextext

%
\IEEEpeerreviewmaketitle

\section{Introduction}
%
%
%
%

\IEEEPARstart{N}{ear}-infrared (NIR) semiconductor lasers are the driving force behind the rapid progress in fiber-optic telecommunication \cite{murphy:2010}.
InP-based materials systems such as \mbox{(GaIn)(AsP)/InP} \cite{phillips:1999} and \mbox{(AlGaIn)As/InP} \cite{zah:1994, chen:1997, higashi:1999, sweeney:2000} have proven to be spectrally suitable choices as active media in lasers emitting in the O- (\SIrange{1.260}{1.360}{\micro\metre}) and C-band (\SIrange{1.530}{1.565}{\micro\metre}).
However, the performance of these devices is affected by Auger losses \cite{meyer:1998} and small band offsets compared to materials systems based on GaAs substrate.
For example, in case of the \mbox{(GaIn)(AsP)/InP} materials system, this results in poor device performances at high temperatures.
Another advantage of GaAs-based technology is the availability of \mbox{(AlGa)As/GaAs} alloys, which for example enable the fabrication of high-quality distributed Bragg reflectors.
While these advantages led to highly efficient lasers based on (AlGaIn)As/GaAs and (GaIn)As/GaAs at wavelengths of \SI{0.808}{\micro\metre} and \SI{0.980}{\micro\metre}, respectively, the design of lasers for the above-mentioned telecommunication bands beyond \SI{1.2}{\micro\metre} has proven to be challenging.
Quantum well (QW) lasers based on (GaIn)(NAs)/GaAs \cite{hoehnsdorf:1999} and Ga(AsSb)/GaAs \cite{yamada:2000} for applications in the O-band were demonstrated but their characteristic properties are challenging in terms of fabrication.
Furthermore, Auger losses cannot be suppressed using these structures.

A potential active medium for semiconductor lasers based on GaAs substrate emitting in the O-band is the (GaIn)As/Ga(AsSb)/GaAs materials system which exhibits a \mbox{type-II} band alignment \cite{peter:1995}.
On the one hand, the \mbox{type-II} band alignment enables a flexible band structure engineering.
On the other hand, \mbox{type-II} heterostructures offer the possibility to suppress Auger losses \cite{zegrya:1995, meyer:1998}.
Thus, \mbox{type-II} semiconductor lasers are a promising candidate for more efficient telecommunication lasers.
Their fabrication, however, has proven to be challenging.
Devices based on double QW heterostructures (QWH) exhibited low pulsed optical output powers of \mbox{P\textsubscript{opt, max} = \SI{140}{\milli\watt}} per facet \cite{klem:2000} and may easily switch from a \mbox{type-II} to a \mbox{type-I} transition under operating conditions \cite{zvonkov:2013}.

A promising approach to optimize optoelectronic devices based on \mbox{type-II} heterostructures is the application of \mbox{``W''-QWHs} as these heterostructures involve a more favorable optical dipole matrix element \cite{chow:2001}.
In order to increase the optical dipole matrix element in these systems, one hole quantum well consisting of Ga(AsSb) is embedded between two electron quantum wells consisting of (GaIn)As in order to increase the confinement function overlap.
Electrical injection lasing with differential efficiencies of \mbox{\texteta\textsubscript{d} = \SIrange{0.12}{0.22}{\watt\per\ampere}}, internal losses of \mbox{\textalpha\textsubscript{i} = \SI{8}{\per\centi\metre}} and an internal quantum efficiency of \mbox{\texteta\textsubscript{i} = \SI{32}{\percent}} at a wavelength of \mbox{\textlambda = \SI{1.2}{\micro\meter}} was achieved using MBE-grown \mbox{``W''-QWHs} \cite{ryu:2002}.
A characteristic temperature of \mbox{T\textsubscript{0} = \SI{68}{\kelvin}} was deduced from temperature dependent measurements in the range of \mbox{T = 10 to \SI{50}{\degreeCelsius}}.
Furthermore, molecular beam epitaxy grown electrical injection lasers emitting at \mbox{\textlambda = \SI{1.3}{\micro\meter}} were reported \cite{chow:2001-2}.

In recent investigations, we demonstrated that the photoluminescence spectra of metal organic vapor phase epitaxy-grown \mbox{``W''-QWHs} can be reproduced by applying a fully microscopic theory.
Furthermore, significant material gain values were predicted based on this experiment-theory comparison \cite{berger:2015}. 
Photomodulated reflectance (PR) spectroscopy was used to confirm the fully microscopic model and to characterize the transitions that contribute to these spectra \cite{gies:2015}.
These transitions include the e1h1 transition between the electron and the hole ground states, the e2h2 transition between the first excited electron and hole states and the e1h3 transition between the electron ground state and the second excited hole state.

Additionally, we have demonstrated a vertical-external-cavity surface-emitting laser with Watt level output powers in transverse multimode operation at a wavelength of \mbox{\textlambda = \SI{1.2}{\micro\meter}} \cite{moeller:2016} as well as fundamental transverse mode operation \cite{moeller:2017}.
The characterization of electrical injection lasers at \mbox{\SI{1.2}{\micro\metre}} yielded low threshold current densities of \mbox{j\textsubscript{th} = \SI{0.4}{\kilo\ampere\per\square\centi\metre}} and \mbox{\SI{0.1}{\kilo\ampere\per\square\centi\metre}} in case of \SI{930}{\micro\metre} and \SI{2070}{\micro\metre} long laser bars, respectively.
Furthermore, high differential efficiencies of \mbox{\texteta\textsubscript{d} = \SI{66}{\percent}}, low internal losses of \mbox{\textalpha\textsubscript{i} = \SI{1.9}{\per\centi\metre}} and optical output powers of  \mbox{P\textsubscript{opt, max} = \SI{1.4}{\watt}} per facet were demonstrated \cite{fuchs:2016}.

As thermal stability is a key feature of  telecommunication lasers, the temperature dependent properties of electrical injection (GaIn)As/Ga(AsSb)/(GaIn)As \mbox{``W''-QWH} lasers are analyzed in the present publication.
For this purpose, theoretical investigations of the gain and band structure properties based on a microscopic approach are carried out in order to provide an in-depth understanding of their charge carrier density-dependent behavior.
These results are used to interpret experimental results obtained from temperature dependent electroluminescence (EL) and laser characteristic measurements.
These are carried out for heat sink temperatures up to \SI{97}{\degreeCelsius} resulting in a comprehensive understanding of the temperature-dependent properties of this \mbox{``W''-QWH} system.

\section{Methods}
\label{methods}

\subsection{Theoretical modeling}
\label{methods-theo}

In order to model the optical response of the W-type QW region and thus simulate the material gain of the given device, we use the well-known Semiconductor Bloch Equations (SBE) that has proven successful in many experiment- theory comparisons \cite{bueckers:2007, bueckers:2008, bueckers:2012} and design studies \cite{thraenhardt:2006, kuehn:2009}.

In this theoretical framework, the laser field is treated classically, while the field-induced optical excitations of the gain material are dealt with on a microscopic level to allow for a highly accurate, quantitatively predictive analysis of the optical properties.
As typical, being interested in the laser behavior around threshold, to keep the numerics feasible, we assume that Coulomb- and Phonon-Scattering events lead to a thermalization of the excited semiconductor material.
In this quasi-equilibrium situation, the charge carriers follow Fermi-Dirac statistics and their detailed dynamics can be neglected.
Higher order scattering terms relevant for the equation of motion for the microscopic polarization that characterizes the system enter on the level of the second Born approximation.
The band structure and wavefunctions relevant for such a semiclassical theory are obtained from an 8x8 Luttinger kp model \cite{chow:1999}.
Additionally, to account for local charge inhomogeneities in the complex heterostructure under consideration, a Schrödinger-Poisson Equation is solved \cite{chow:1999, haug:2009}.
Intrinsic statistical alloy fluctuations, layer variations and interface roughness get involved by convolving the optical gain spectra computed on basis of all these model assumptions with a Gaussian.

\subsection{Experimental methods}
\label{methods-exp}

All devices investigated in the present work are fabricated using metal organic vapor phase epitaxy \cite{fuchs:2017} and processed as gain-guided broad-area edge-emitting lasers with stripe widths of \SI{100}{\micro\metre} \cite{fuchs:2016}.

Device characterization is carried out in a p-side up geometry on a copper heat sink allowing for a temperature variation between \SI{10}{\degreeCelsius} and \SI{100}{\degreeCelsius} using a Peltier device.
All measurements are carried out under pulsed operating conditions using a pulse length of \SI{400}{\nano\second} and a repetition rate of \SI{10}{\kilo\hertz} resulting in a duty cycle of \SI{0.4}{\percent}.
Spectral measurements are performed by recording electroluminescence spectra using a grating monochromator in combination with a lock-in amplifier in case of measurements below laser threshold and using a optical spectrum analyzer in case of measurements above laser threshold.
Furthermore, laser characteristics are recorded using a large-area germanium photodetector. 

 

\section{Theoretical investigation}
\label{sec:theory}

The following theoretical investigation outlines the confinement potentials as well as material gain properties of (GaIn)As/Ga(AsSb)/(GaIn)As \mbox{``W''-QWHs} emitting at approximately \SI{1.3}{\micro\metre}.
This emission wavelength is obtained assuming (GaIn)As and Ga(AsSb) layer thicknesses of \SI{4}{\nano\metre} in combination with In- and Sb-concentrations of \SI{28}{\percent}.
This \mbox{``W''-QWHs} design results in a low-density luminescence emission wavelength of approximately \SI{1345}{\nano\metre}.

\subsection{Carrier density-dependence of the confinement potentials}
\label{sec:theory-bandstructurecarrierensity}

\begin{figure}[!ht]
\centering
	\includegraphics[width = 8.5cm]{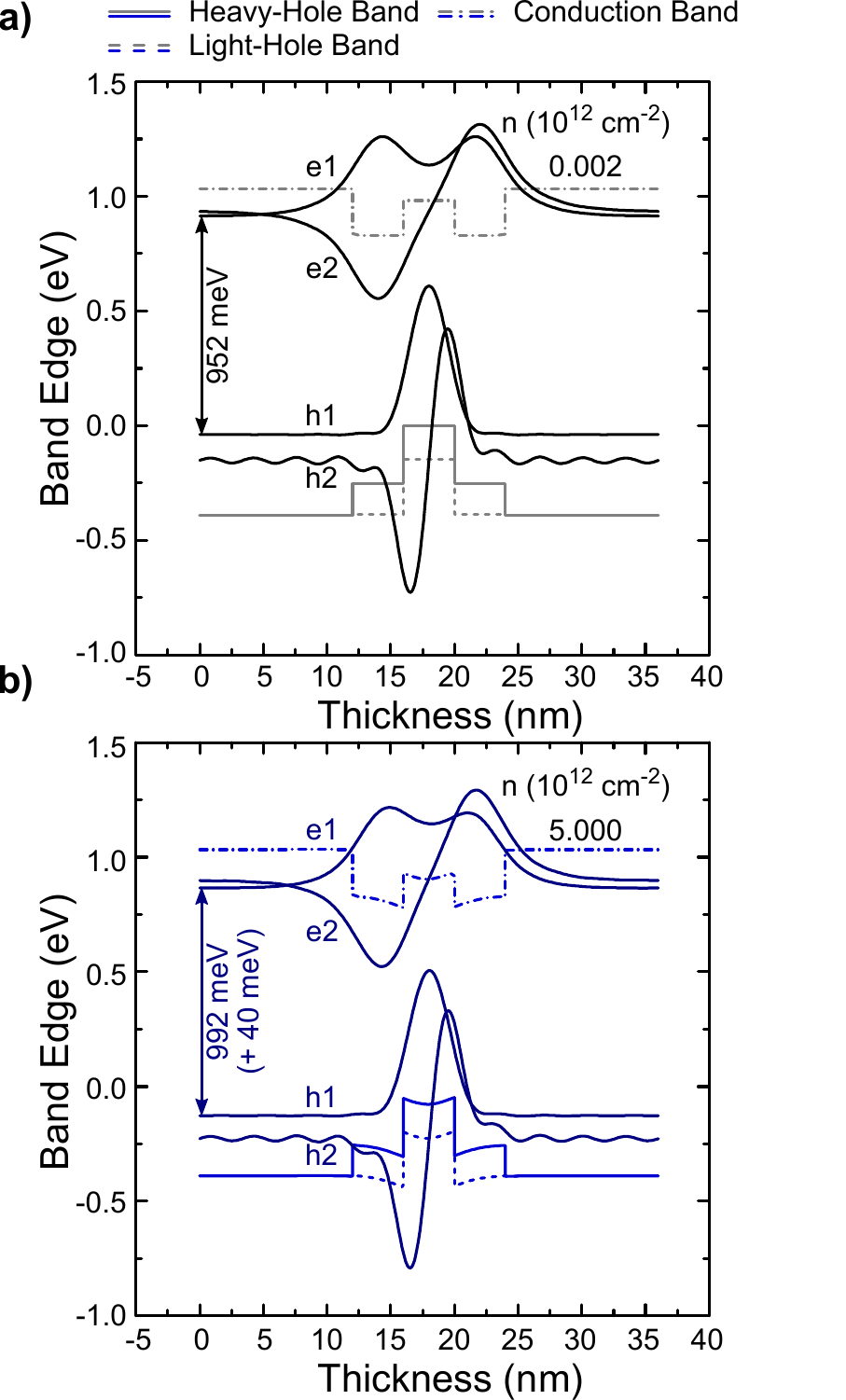}
    \caption{Theoretically calculated confinement potentials and functions for the \mbox{``W''-QWH} at charge carrier densities of a)~\SI{0.002e12}{\per\square\centi\metre} and b)~\SI{5.000e12}{\per\square\centi\metre}.}
  \label{fig:TheoBandStructureCarrierDensity}
\end{figure}

While material gain calculations for (GaIn)As/Ga(AsSb)/(GaIn)As \mbox{``W''-QWHs} are available in the literature \cite{chow:2001, chow:2001-2, berger:2015}, the confinement potentials of these heterostructures are rarely discussed in more detail.
Thus, the starting point of the present work is an investigation of the charge-carrier density-dependence of the confinement potentials as well as the corresponding confinement functions.
The confinement potentials as well as the confinement functions of the two energetically lowest electron and hole states for the above-mentioned structure is shown in Fig.~\ref{fig:TheoBandStructureCarrierDensity}\,a) assuming a low charge carrier densities of \SI{0.002e12}{\per\square\centi\metre}.
At this carrier density, the band edges are box-like and both electron states (i.e. e1 \& e2) are almost degenerate with an energetic separation of only \SI{21}{\milli\electronvolt}.
Furthermore, an energetic separation of \SI{948}{\milli\electronvolt} between the electron and hole ground states (i.e. e1 \& h1) is observed.

These properties change when the charge carrier density in the \mbox{``W''-QWH} is increased to \SI{5.000e12}{\per\square\centi\metre}, which we consider a typical charge carrier density at which laser operation is achieved in semiconductor diode lasers.
The confinement potentials are strongly distorted due to the spatial separation of electrons and holes.
The spatial separation of positive and negative charge carriers results in an internal electrical field caused by the Coulomb interaction between both carrier species.
Consequently, the overlap between the ground state confinement functions as well as the energetic separation between the electron states is increased to \SI{32}{\milli\electronvolt} as shown in Fig.~\ref{fig:TheoBandStructureCarrierDensity}\,b).
The energetic separation between both ground state levels is also is blue shifted by \SI{37}{\milli\electronvolt} up to a total of \SI{985}{\milli\electronvolt}.

\subsection{Carrier density dependence of the material gain}
\label{sec:theory-gaincarrierensity}

\begin{figure}[!ht]
\centering
	\includegraphics[width = 8.5cm]{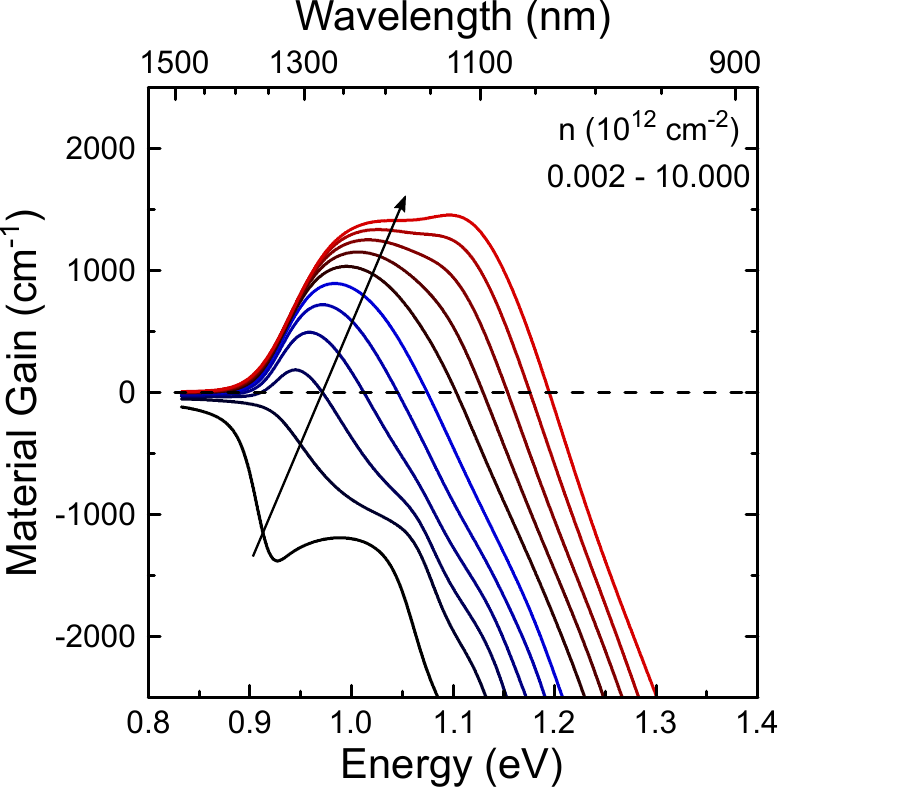}
    \caption{Theoretically calculated material gain spectra for the \mbox{``W''-QWH} at charge carrier densities of between \SI{0.002e12}{\per\square\centi\metre} and \SI{10.000e12}{\per\square\centi\metre} in steps of \SI{1.000e12}{\per\square\centi\metre}.}
  \label{fig:TheoGainCarrierDensity}
\end{figure}

In order to determine the implications of the above-mentioned behavior for semiconductor laser applications, gain spectra are calculated for charge carrier densities between \SI{0.002e12}{\per\square\centi\metre} and \SI{10.000e12}{\per\square\centi\metre} assuming an inhomogeneous broadening of \SI{20}{\milli\electronvolt} as shown in Fig.~\ref{fig:TheoGainCarrierDensity}.
While these gain spectra show similar characteristics compared to type-I heterostructures reaching transparency and building up significant material gain values as the charge carrier density is increased, two important differences are observed.
Firstly, the material gain of type-I heterostructures is expected to be red shifted due to band gap renormalization as the charge carrier density is increased.
However, \mbox{type-II} heterostructures exhibit a characteristic blue shift primarily caused by the above-mentioned band bending, which was also outlined in previous studies \cite{chow:2001, chow:2001-2, berger:2015}.
This blue shift amounts to approximately \SI{52}{\nano\metre} in case the presently investigated \mbox{``W''-QWH} corresponding to a shift of the gain peak from \SI{1313}{\nano\metre} at \SI{2.000e12}{\per\square\centi\metre} to \SI{1261}{\nano\metre} at \SI{5.000e12}{\per\square\centi\metre}.
Secondly, the material gain spectra are intitially dominated by the ground state transition between the e1 and h1 states.
However, as the charge carrier density is increased above \SI{5.000e12}{\per\square\centi\metre}, the transition between the e2 and h2 states slowly starts to dominate the gain spectra.
Due to the small energetic separation between the electron states, the energetic separation between the hole states of \SI{113}{\milli\electronvolt} at \SI{5.000e12}{\per\square\centi\metre} largely defines the separation between both transitions.
Thus, an overlap between the e1h1 and e2h2  contributions is observed resulting in an almost flat gain spectrum over a range of approximately \SI{160}{\nano\metre} at \SI{10.000e12}{\per\square\centi\metre}.
This result highlights the importance of a thorough theoretical design of these \mbox{type-II} heterostructures for actual device applications, because such a behavior might result in semiconductor lasers switching to higher order transitions under certain operating conditions.
However, these properties must also be considered as an unique opportunity for the design of novel semiconductor devices such as semiconductor optical amplifiers, where broad and flat gain spectra are highly beneficial.

\subsection{Investigation of design-parameters for tailoring the material gain}
\label{sec:theory-designmaterialgain}

While these results seem promising, it is also important to show that the charge carrier density-dependence can be tailored in order to be able to adapt the active regions for specific device applications.
Therefore, the gain calculations are repeated for different \mbox{``W''-QWH} designs, where the (GaIn)As and Ga(AsSb) thicknesses are systematically varied between \SI{4}{\nano\metre} and \SI{8}{\nano\metre} while keeping either the (GaIn)As or Ga(AsSb) thickness constant and maintaining constant In and Sb concentrations.
Furthermore, the \mbox{``W''-QWH} is assumed to be symmetric in all calculations, i.e. both (GaIn)As quantum wells are assumed to exhibit the same thickness.
Three important properties of the flat gain spectra are monitored in this investigation:
 
\begin{enumerate}
\item The charge carrier density required to reach flat gain spectra n\textsubscript{flat}
\item The energetic witdh \textDelta\textlambda\textsubscript{w} in nm (or \textDelta E\textsubscript{w} in meV) of the flat gain region
\item The plateau gain value g\textsubscript{flat}
\end{enumerate}

The investigation shows that increasing layer thicknesses of either (GaIn)As or Ga(AsSb) results in flat gain spectra with smaller spectral widths \textlambda\textsubscript{w} and  a lower gain plateau g\textsubscript{flat} being obtained at lower charge carrier densities n\textsubscript{flat}. Despite these comparably small changes, it is possible to modify all above-mentioned properties by a factor of 2-3 compared to the initial structure with \SI{4}{\nano\metre} thick (GaIn)As and Ga(AsSb) QWs. These observations are explained by multiple underlying effects including:

\begin{itemize}
\item Reduced energetic separation between confined sub-band states as layer thicknesses are increased resulting in higher occupation probabilities of excited states
\item Lower confinement function overlaps between electron and hole sub-band confinement functions
\item Smaller impact of of band bending effects due to larger active regions
\item The changing density of states resulting from the larger layer thicknesses
\end{itemize}

\section{Experimental findings}
\label{sec:exp}

In addition to these theoretical investigations, a series of laser samples emitting at approximately \SI{1.2}{\micro\metre} is investigated experimentally.
The sample series was chosen in such a way that the influence of the charge carrier density per active ``W''-QWH can be investigated.
In order to do so, sample A and B were fabricated using the same growth conditions.
However, the active region of sample A only includes a single ``W''-QWH, while sample B includes two active ``W''-QWHs.
Additionally, the influence of interface modifications in the form of thin GaP layers surrounding the outer interfaces of the ``W''-QWH are investigated.
This modification aims at improving electron-hole confinement function overlap by introducing step-like barriers at the edge of the ``W''-QWH.
Relevant material properties of these samples are summarized in Tab.~\ref{tab:samples}.

\begin{table*}[!ht]
\centering
	\caption{Summary of material properties of samples A, B and C including the cavity length (L), contact width (W), and emission wavelength (\textlambda) at a heatsink temperature of \SI{20}{\degreeCelsius}.}
  	\label{tab:samples}
	\begin{tabular}{ccccccccccccc}
  	\hline \hline 
	Sample & Design & d\textsubscript{(GaIn)As} (nm) & c\textsubscript{In} & d\textsubscript{Ga(AsSb)} (nm) & c\textsubscript{Sb} & L (\si{\micro\metre})& W (\si{\micro\metre}) & \textlambda~(\si{\nano\metre}) \\
	\hline
	A & Single ``W''-QWH & 6 & 0.2 & 4 & 0.2 & 930 & 100 & 1155 \\
	B & Double ``W''-QWH & 6 & 0.2 & 4 & 0.2 & 990 & 100 & 1186 \\
	C & Single ``W''-QWH with GaP interlayers & 6 & 0.2 & 4 & 0.2 & 970 & 100 & 1176 \\
	\hline\hline
\end{tabular}
\end{table*}

\subsection{Temperature-dependent spectral properties of (GaIn)As/Ga(AsSb)/(GaIn)As ``W''-quantum well lasers}
\label{sec:exp-spectral}

\begin{figure}[!ht]
\centering
	\includegraphics[width = 8.5cm]{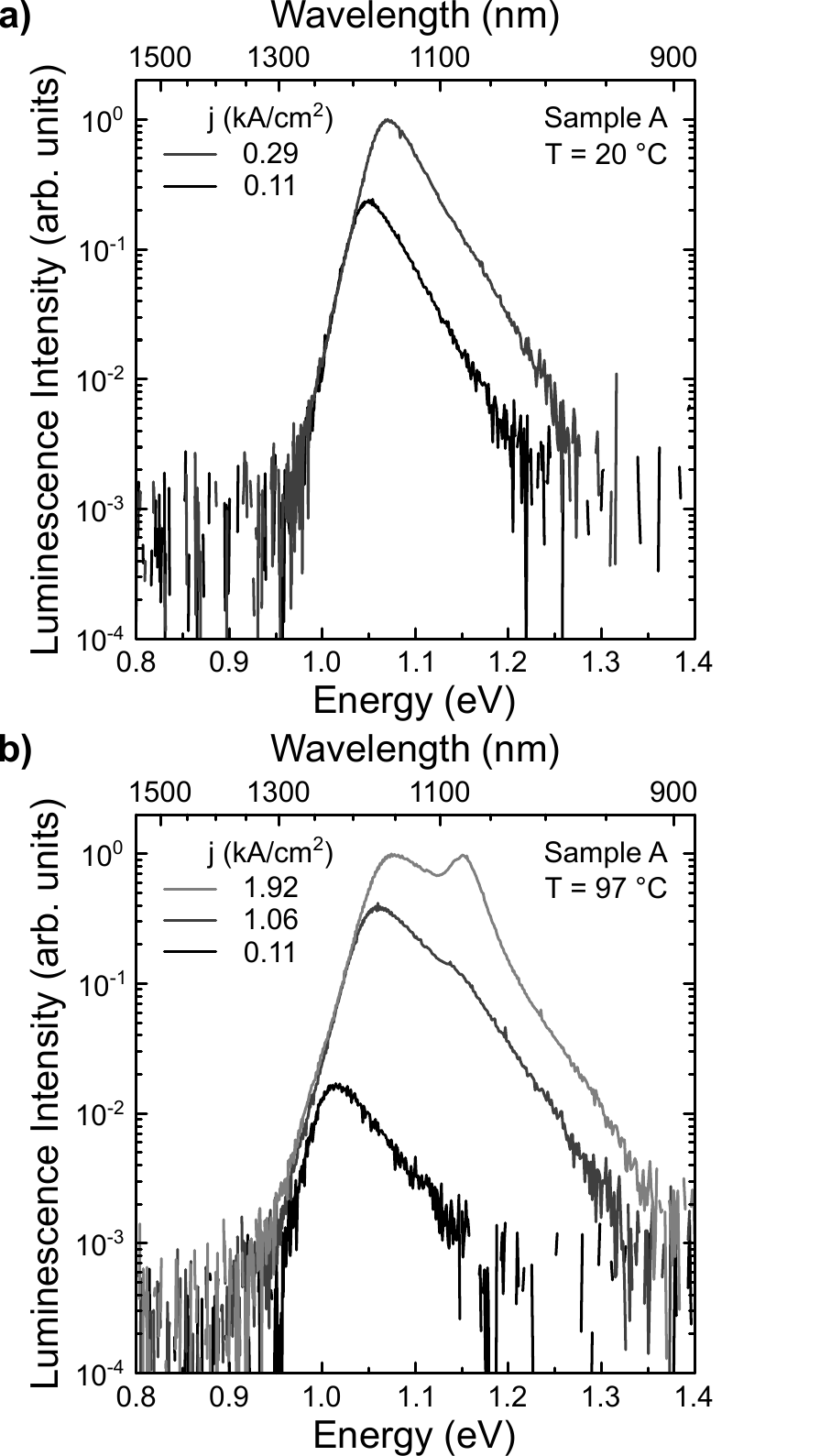}
    \caption{Electroluminescence spectra below laser threshold of sample A at a) \SI{20}{\degreeCelsius} and b) \SI{97}{\degreeCelsius} for different current densities.}
  \label{fig:ExpSpectra}
\end{figure}

The present work aims at an in-depth investigation of temperature-dependent properties of (GaIn)As/Ga(AsSb)/(GaIn)As ``W''-QWH lasers complementing previously published room temperature data \cite{fuchs:2016}.
In order to do so, the respective electroluminescence spectra below laser threshold are compared at different temperatures of \SI{20}{\degreeCelsius} and \SI{97}{\degreeCelsius} as shown in Fig.~\ref{fig:ExpSpectra}.
The measurements verify that sample exhibits a significant blue shift as the current density is increased from \SI[per-mode=symbol]{0.11}{\kilo\ampere\per\square\centi\metre} to \SI[per-mode=symbol]{0.29}{\kilo\ampere\per\square\centi\metre} at a temperature of \SI{20}{\degreeCelsius} \cite{chow:2001, chow:2001-2, berger:2015}.
However, as the temperature is increased to \SI{97}{\degreeCelsius} resulting in an increased threshold current density, the sample exhibits a significant side peak which eventually starts to exceed the fundamental \mbox{type-II} transition.
As a consequence, laser emission is no longer based on the fundamental \mbox{type-II} transition, but based on this excited \mbox{type-II} transition, which is identified as e2h2 transition using the theoretical model outlined above.
This finding indicates that the modal gain contributed by the fundamental \mbox{type-II} transition eventually saturates prior to exceeding the total loss at this elevated temperature.
Excess charge carriers start to populate the higher order states until the modal gain contributed by this transition eventually exceeds the losses and laser operation is observed.

These observations highlight the importance of a careful microscopic design of these ``W''-QWH active regions in order to achieve stable operation across a desired temperature range.
However, as laser operation based on both of these \mbox{type-II} transitions is observed in case of this sample, these findings are a first indication for the presence of the broad material gain spectra described above.

\subsection{Temperature-dependent laser properties of (GaIn)As/Ga(AsSb)/(GaIn)As ``W''-quantum well lasers}
\label{sec:exp-laser}

\begin{figure}[!ht]
\centering
	\includegraphics[width = 8.5cm]{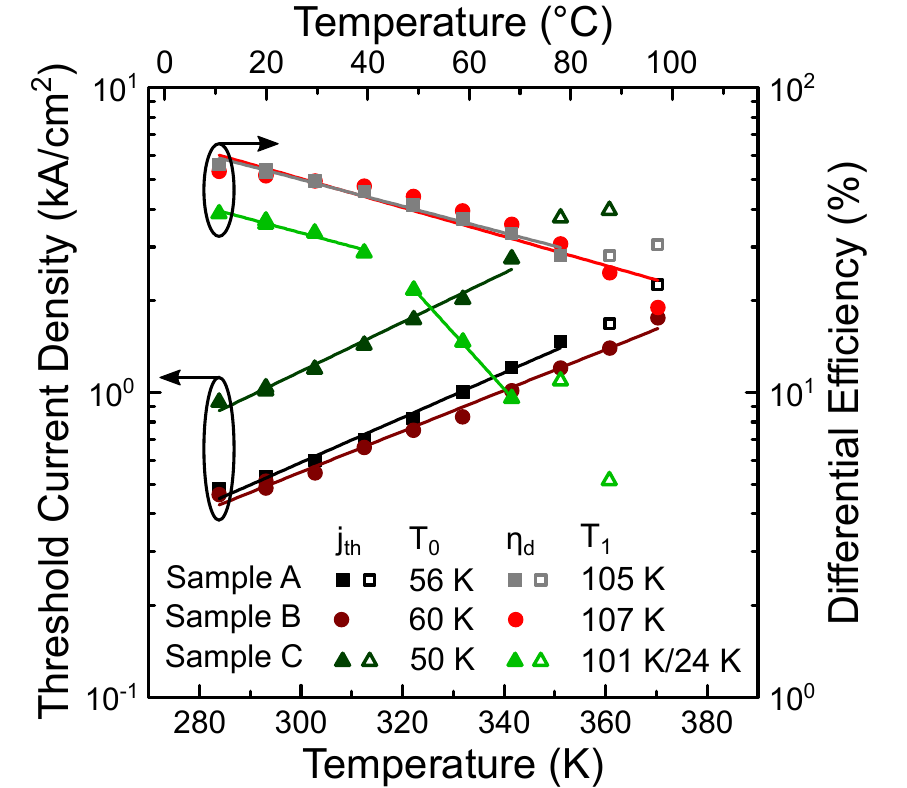}
    \caption{Temperature-dependence of the threshold current density and the differential efficiency of sample A, B, and C. Open symbols indicate operation based on an excited \mbox{type-II} transition. Exponential fits yield characteristic temperatures of \mbox{T\textsubscript{0} = (56 $\pm$ 2)\,\si{\kelvin}} and \mbox{T\textsubscript{1} = (105 $\pm$ 6)\,\si{\kelvin}} for sample A and \mbox{T\textsubscript{0} = (60 $\pm$ 2)\,\si{\kelvin}} and \mbox{T\textsubscript{1} = (107 $\pm$ 12)\,\si{\kelvin}} for sample B. While sample C also shows a single slope behavior in case of \mbox{T\textsubscript{0} = (50 $\pm$ 3)\,\si{\kelvin}}, different characteristic temperatures \mbox{T\textsubscript{1} = (101 $\pm$ 13)\,\si{\kelvin}} and \mbox{T\textsubscript{1} = (24 $\pm$ 1)\,\si{\kelvin}} are observed below and above \SI{50}{\degreeCelsius}, respectively.}
  \label{fig:ExpT0T1}
\end{figure}

In order to investigate the feasibility of these ``W''-QWHs for laser applications, laser characteristics are recorded for temperatures between \SI{11}{\degreeCelsius} and \SI{97}{\degreeCelsius}.
The temperature-dependent behavior of the threshold current density as well as differential efficiency are summarized in Fig.~\ref{fig:ExpT0T1}, where filled symbols indicate operation based on the fundamental and open symbols indicate operation based on the excited \mbox{type-II} transition.
Since operation based on the fundamental transition is desired for most laser applications, the following analysis excludes data points above the switching temperature.
Samples A and B exhibit similar properties across the entire temperature range for the threshold current density as well as the differential efficiency.
Exponential fits yield characteristic temperatures of \mbox{T\textsubscript{0} = (56 $\pm$ 2)\,\si{\kelvin}} and \mbox{T\textsubscript{1} = (105 $\pm$ 6)\,\si{\kelvin}} for sample A and \mbox{T\textsubscript{0} = (60 $\pm$ 2)\,\si{\kelvin}} and \mbox{T\textsubscript{1} = (107 $\pm$ 12)\,\si{\kelvin}} for sample B.
Thus, these characteristic temperatures appear to be independent of the charge carrier density per active ``W''-QWH.
However, the switching to a higher order transition is fully suppressed in this temperature range by the introduction of a second ``W''-QWH as previously shown in the literature \cite{fuchs:2017-2}.

In case of sample C, the modification of the active region using GaP interlayers resulted in a deterioration of the device properties.
A significant increase in threshold current density and a decrease of its temperature-dependence (\mbox{T\textsubscript{0} = (50 $\pm$ 3)\,\si{\kelvin}}) is observed.
Furthermore, differential efficiencies are decreased and a two slope exponential model is required to characterize the temperature-dependence of the differential efficiency.
This analysis yields \mbox{T\textsubscript{1} = (101 $\pm$ 13)\,\si{\kelvin}} and \mbox{T\textsubscript{1} = (24 $\pm$ 1)\,\si{\kelvin}} for temperatures below and above \SI{50}{\degreeCelsius}, respectively.
Further investigations are required in order to explain these findings due to the complex nature of the microscopic changes introduced in this structure.
Possible root causes include among others the introduction of loss or leakage channels through interface states, dissimilar injection efficiencies for both charge carrier species as well as phosphorus segregation resulting in an asymmetry of the ``W''-QWHs.

\begin{figure}[!ht]
\centering
	\includegraphics[width = 8.5cm]{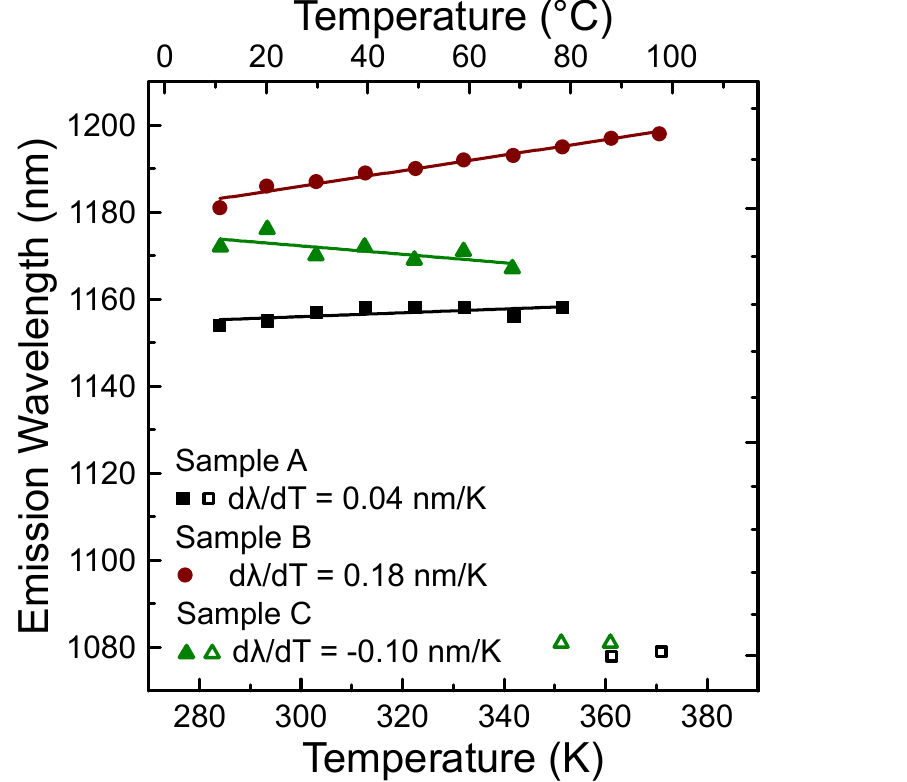}
    \caption{Temperature-dependence of the laser emission wavelength of sample A, B, and C. Linear fits yield shift rates of (0.04 $\pm$ 0.02)\,\si[per-mode=symbol]{\nano\metre\per\kelvin}, (0.17 $\pm$ 0.01)\,\si[per-mode=symbol]{\nano\metre\per\kelvin}, and (-0.10 $\pm$ 0.04)\,\si[per-mode=symbol]{\nano\metre\per\kelvin}, respectively.}
  \label{fig:ExpWavelength}
\end{figure}

In case of regular \mbox{type-I} QWH lasers, these findings could be interpreted as poor temperature stability.
In \mbox{type-II} ``W''-QWH lasers, however, low T\textsubscript{0} values are particularly interesting due to the interplay between the charge carrier density-induced blue shift below threshold and the temperature-induced red shift of the emission wavelength.
In order to investigate this effect, laser spectra above threshold are recorded together with each laser characteristic used to determine the characteristic temperatures.
Thermal shift rates of the emission wavelength are determined for samples A, B and C up to  their respective temperature, where laser emission switches from the fundamental to the excited transition.
The present devices exhibit thermal shift rates of (0.04 $\pm$ 0.02)\,\si[per-mode=symbol]{\nano\metre\per\kelvin} (sample A), (0.17 $\pm$ 0.01)\,\si[per-mode=symbol]{\nano\metre\per\kelvin} (sample B), and (-0.10 $\pm$ 0.04)\,\si[per-mode=symbol]{\nano\metre\per\kelvin} (sample C) as shown in Fig.~\ref{fig:ExpWavelength}.
The difference in shift rates between these samples can be explained by the difference in microscopic as well as device design since growth conditions were kept constant across devices.
The addition of a second ``W''-QWH to the active region in sample B results in a distribution of carriers into both ``W''-QWHs and as a result, the resulting blue shift is decreased and outweighed by the thermal red shift.
In case of sample C, the addition of GaP interlayers appears to deteriorate the overall device performance and significantly lowers T\textsubscript{0} resulting in a stronger charge carrier density-induced blue shift as the temperature is increased.

These findings can be considered as unique property of \mbox{type-II} heterostructure lasers and strongly differentiate them from regular type-I heterostructure lasers, where a thermal shift rate of approximately \SI[per-mode=symbol]{0.4}{\nano\metre\per\kelvin} is expected in this temperature range.
These findings open up novel device applications, where \mbox{type-II} ``W''-QWHs are tailored in such a way that their thermal shift rate of the emission wavelength matches the respective device needs.
Examples include distributed feedback lasers, distributed Bragg reflector lasers or vertical-cavity surface-emitting lasers, where the thermal shift rate of the active region is matched to the grating or reflector in order to improve the temperature stability of these devices.
Furthermore, the interplay of resistive heating and the charge carrier density-induced blue shift in devices operated under continuous wave operating conditions is expected to also result in smaller thermal shift of the emission wavelength.
Classical design rules, which demand high characteristic temperatures to lower the power dissipation of devices, can be circumvented since the demand for external cooling can be significantly decreased due to the more temperature stable emission wavelength.

\section{Conclusion}
In conclusion, the theoretical and experimental investigation of \mbox{type-II} (GaIn)As/Ga(AsSb)/(GaIn)As ``W''-quantum well heterostructures indicate a significant application potential of these heterostructures.
On the one hand, the microscopic theory predicts the existence of flat and broad gain spectra with spectral widths exceeding \SI{100}{\nano\metre}.
First indications for the existence of these broad gain spectra are being observed in temperature-dependent electroluminescence measurements.
On the other hand, experimental findings show that it is possible to use the well-understood charge carrier density-induced blue shift in these \mbox{type-II} ``W''-quantum well heterostructures to compensate the temperature-induced red shift of the emission wavelength in semiconductor lasers.
Thus, it is possible to realize devices with thermal shift rates of the emission wavelength ranging from typical \mbox{type-I} values of \SI[per-mode=symbol]{0.4}{\nano\metre\per\kelvin} down to negative values of \SI[per-mode=symbol]{-0.1}{\nano\metre\per\kelvin} and potentially even further.
These findings could result in significant improvements of devices such as distributed feedback, distributed Bragg reflector and vertical-external-cavity surface-emitting lasers as well as semiconductor optical amplifiers and frequency combs.
Future research should focus on an in-depth understanding of loss and leakage mechanisms in these devices in order to further tailor the device properties for specific applications.



%



\section*{Acknowledgment}

The authors gratefully acknowledge the funding provided by Deutsche Forschungsgemeinschaft (DFG) in the framework of Sonderforschungsbereich 1083 - Structure and Dynamic of Internal Interfaces - and the framework of the Research Training Group 1782 -
 Functionalization of Semiconductors. The Tucson work was supported by the Air Force Office of Scientific Research under award number FA9550-17-1-0246.

\ifCLASSOPTIONcaptionsoff
  \newpage
\fi

\end{document}